\documentstyle[epsf,twocolumn,prl,aps]{revtex}

\newcommand{\Nel}{N_{\rm el}}

\begin{document}
\draft

 \twocolumn[\hsize\textwidth\columnwidth\hsize\csname @twocolumnfalse\endcsname

\title{Ab Initio Quantum Chemistry using the Density Matrix
Renormalization Group}
\author{ Steven R.\ White$^1$ and Richard L. Martin$^2$}
\address{ 
$^1$Department of Physics and Astronomy,
University of California,
Irvine, CA 92697
}
\address{ 
$^2$Theoretical Division, MSB268, Los Alamos National Laboratory, Los Alamos, NM 87545
}
\date{\today}
\maketitle
\begin{abstract}
\noindent 

In this paper we describe how the density matrix renormalization 
group (DMRG) can be used for quantum chemical calculations for
molecules, as an alternative to traditional methods, such as
configuration interaction or coupled cluster approaches.
As a demonstration of the potential of this approach, we
present results for
the H$_2$O molecule in a standard gaussian basis.
Results for the total energy of the system 
compare favorably with
the best traditional quantum chemical methods.

\end{abstract}

 ]

Since its development in 1992\cite{dmrg}, the density matrix renormalization
group (DMRG) has become one of the most widely used numerical
techniques for simulations of one dimensional quantum lattice
systems. For systems with short ranged interactions, the 
calculation time for DMRG grows only linearly with
the length of the system, while the errors usually decrease
exponentially with the calculation time\cite{details}.
Consequently, very high accuracy results are possible even on
very large systems. 

Most applications of DMRG have been to lattice models for
strongly correlated systems.
Recently, Fano, Ortolani, and Ziosi applied DMRG 
to a Pariser-Parr-Pople (PPP) Hamiltonian
for a cyclic polyene\cite{fano}. 
The PPP model is more realistic than many models in that
long range coulomb interactions are included.  Fano, et. al.
found that DMRG compared quite favorably to coupled cluster approximations. 
Here, we go a step further: we consider the application of DMRG to the 
fully {\it ab initio} determination of the electronic structure of atoms and
molecules. The successful adaptation of DMRG to this field
could potentially open up a wide range of improved calculational
techniques, characterized by high accuracy and improved scaling
of calculation time with system size. As a first step in this
direction, we show here that DMRG can be successfully used to
obtain very accurate 
many-body solutions for small molecules.

We will use DMRG within the conventional 
quantum chemical framework of a finite basis set with
non-orthogonal basis functions made from products of gaussian
radial functions and 
Cartesian harmonics centered on each atom.
The initial step
of the calculation is a standard Hartree Fock (HF) calculation 
in which a Hamiltonian is produced within the orthogonal HF basis.
DMRG is then used as a procedure for including correlations
beyond HF, much as the configuration interaction (CI) or
coupled cluster methods are used.

Within the HF basis, the Hamiltonian is in principle
no different from other model Hamiltonians which have been
studied using DMRG. It can be written as
\begin{equation}
H = \sum_{ij\sigma} T_{ij} c^\dagger_{i\sigma}c_{j\sigma}
+ \frac{1}{2} \sum_{ijkl\sigma\sigma'} V_{ijkl} 
c^\dagger_{i\sigma}c^\dagger_{j\sigma'}c_{k\sigma'}c_{l\sigma} .
\end{equation}
Here $T_{ij}$ contains the electron kinetic energy and the
Coulomb interaction between the electrons and the nuclei, while
$V_{ijkl}$ describes the electron-electron Coulomb interaction.
The most important difference in this Hamiltonian from model
Hamiltonians is the large number of interaction terms
$V_{ijkl}$: $N^4$, where $N$ is the number of basis functions or
orbitals. 
The number of electrons, $\Nel$, is less important in
a DMRG calculation.
The large number of terms makes standard DMRG programs very
inefficient, and below we describe procedures for improving the
efficiency of the treatment of these terms\cite{notoned}.

In our approach, an ordering of the orbitals is
chosen, and each orbital is treated as a ``site'' 
in a one-dimensional lattice. Since this arrangement is
artificial, the Hamiltonian is long-ranged. The orbitals can be
sorted according to various criteria. We have found that sorting
them in order to minimize strong interactions between widely
separated orbitals is probably best, but they can also be
arranged by HF orbital energy. 

Once this ordering is
chosen, a standard DMRG finite-system algorithm can be
used\cite{dmrg}.
In this procedure, collections of orbitals are represented
as ``blocks''. The properties of a block are defined by listing
the many-body states of the block and by storing matrices
representing operators acting on that collection of orbitals.
The representation is approximate, since not all of the
many-body states are retained. For example, if the block
happened to represent the $\Nel/2$ occupied HF orbitals, then a
reasonable set of states to represent that block would consist
of one ``filled'' state with $\Nel$ electrons, $\Nel$ one-hole 
states with $\Nel-1$ electrons, and $(\Nel^2-\Nel)/2$ two-hole states
with $\Nel-2$ electrons. Assuming the one- and two-particle states were
represented in the ``unoccupied'' block, this set of states would allow all
singly and doubly excited configurations to be formed, but would
leave out all higher excitations. DMRG is an iterative
procedure, in which at each step there are two blocks,
with all the orbitals belonging to one of the two
blocks\cite{twosite}.
Iterations, or sweeps, involve transfering orbitals one at a time from the
right to the left block, until the right block has only one
orbital, and then reversing the direction. At each step, a new
set of states is chosen to represent the block. 
The number of states $m$ kept per block controls the accuracy of
the calculation, as well as the storage and computation time.

However, rather than one having to choose the many-body states 
which describe a block, the states are chosen in an optimal
way by DMRG as the eigenstates of a many-particle density
matrix. This procedure is somewhat related to the use of a
single particle density
matrix to choose natural orbitals in quantum chemistry. However, 
here the many particle states are much more complicated---too
complicated, for a system of reasonable size, to represent in
terms of a single particle basis. Instead, the states are 
described in terms of the matrix elements of various operators between
these states. The complicated form of the many particle basis
allows much more rapid convergence in the number of states $m$ than
in a configuration expansion.

The operators which describe a block are chosen in order to be
able to generate the Hamiltonian operator for the system.
For example, in order to construct the kinetic energy, we must
keep matrix elements for the operators $c^\dagger_{i\sigma}$,
for all orbitals $i$ in the block. These operators allow us to
construct terms $c^\dagger_{i\sigma}c_{j\sigma}$ where $j$ is
not in the block.  In addition, we must keep
matrices for $c^\dagger_{i\sigma}c_{j\sigma}$ if both $i$ and
$j$ are in the block. 
Note that one cannot avoid storing a matrix for $AB$ simply
because one has stored matrices for $A$ and $B$: the incomplete
nature of the basis means that the matrix for $AB$ is not the
product of the matrices for $A$ and $B$.
In order to describe the Coulomb
interaction, it appears that $o(N^4)$ operators of the form
$c^\dagger_{i\sigma}c^\dagger_{j\sigma'}c_{k\sigma'}c_{l\sigma}$,
must be kept, where $i$, $j$, $k$, and $l$ are all in the block.
In addition, $o(N^3)$ additional operators are needed to
construct terms when some of the $ijkl$ are not in the block.

Although it appears that $o(N^4)$ operator matrices must be
stored per block, a completely standard DMRG ``trick'' reduces this
number to $o(N^3)$. The trick is to sum terms together into a
single block Hamiltonian matrix once all of the parts of the
term are in the block. Hence there is no need to store terms of the
form $c^\dagger_{i\sigma}c^\dagger_{j\sigma'}c_{k\sigma'}c_{l\sigma}$;
these terms are multiplied by  $V_{ijkl}$ and summed into $H$.
In a typical model Hamiltonian, this tricks reduces the number
of operators stored per block from $o(N)$ to $o(1)$. Here, we
still have $o(N^3)$ operators. Since $o(N)$ blocks must be
stored, the storage is $o(N^4 m^2)$.

Additional improvements can be made by combining other
operators, as was first done by Xiang in adapting DMRG to
momentum space calculations\cite{xiang}. There are $o(N^3)$
operators with three $c$ and $c^\dagger$ operators, used to
construct the Coulomb interaction. These can be largely
eliminated by constructing complementary operators like
\begin{equation}
O_{i\sigma} = \sum_{jkl\sigma'} V_{ijkl}
c^\dagger_{j\sigma'}c_{k\sigma'}c_{l\sigma}
\end{equation}
The corresponding parts of the Coulomb interaction can be
constructed as
\begin{equation}
\sum_{i\sigma} c^\dagger_{i\sigma} O_{i\sigma}.
\end{equation}
This trick reduces $o(N^3)$ operators to $o(N)$.

At this point, the dominant terms remaining are $o(N^2)$ operators with
two $c$ and $c^\dagger$s. The total storage for these, $o(N^3 m^2)$,
is now manageable.
Additional complementary operators can reduce computation time,
however. The dominant part of a DMRG calculation is the
iterative diagonalization of the Hamiltonian of the system,
which is done once per step, or $\sim N$ times per sweep.
The dominant part of this is the multiplication of a vector by
Hamiltonian terms of the form
\begin{equation}
\sum_{ij\in L}
\sum_{kl\in R} V_{ijkl}
[c^\dagger_{i}c^\dagger_{j}][c_{k}c_{l}]
\end{equation}
plus other combinations where, for example, $i$ and $l$ belong
to the left block $L$, $j$ and $k$ to the right block $R$. Here
$[]$ denote the matrix for the corresponding operator. There are
$o(N^4)$ such terms; multiplication of a vector by these terms
requires $o(N^4 m^3)$ operations. To reduce this computation
time, we construct complementary operators
\begin{equation}
O_{ij}^R = \sum_{kl\in R} V_{ijkl} [c_{k}c_{l}]
\qquad \forall ij \in L
\end{equation}
plus other two operator combinations corresponding to other
orbitals being in the left block.
Constructing these operators at each step requires $o(N^4 m^2)$ 
operations. However, this is not necessary: one can save these
operators from the previous step, transform them to the current
basis, and add in the additional terms coming from the new site
being added to the block, in $o(N^2 m^3) + o(N^3 m^2)$ operations.
Using the complementary operators, corresponding Hamiltonian terms
become
\begin{equation}
\sum_{ij \in L} [c^\dagger_{i}c^\dagger_{j}]O_{ij}^R
\end{equation}
plus similar terms. The calculation time to multiply a vector by
these terms is $o(N^2 m^3)$, giving $o(N^3 m^3)$ per DMRG sweep.

After all of these optimizations are used, the final calculation
time for the whole calculation is $o(N^3 m^3) + o(N^4 m^2)$. 
The final storage is $o(N^3 m^2)$, but only $o(N^2 m^2)$ needs
to be in RAM; the rest can be on disk with little cost in
calculation time.
The time for the initial Hartree Fock calculation is $o(N^4)$,
which is neglible in comparison. In the test calculations below, 
the number of states kept per block $m$ is typically a few
hundred, and $N=25$. However, in cases such as linear chain
molecules, we expect to be able to hold $m$ constant as the
the length of the system increases, so that eventually $N \sim m$.
If high accuracy was not required, one could have $N >> m$.
Note that even $m=1$ gives results slightly more accurate than
Hartree Fock. If a localized basis, rather than the Hartree Fock
basis, were used in a large molecule,
so that many coefficients $V_{ijkl}$ were neglible, then 
the calculation time could potentially be reduced to 
$o(N^2 m^3)$. If, in addition, the long range part of the
Coulomb interaction could be neglected, 
as is done in many model
Hamiltonians, 
or closely approximated 
with a multipole expansion, 
the calculation time would be $o(N m^3)$. 
However, in the current {\it ab initio} calculations we are far
from this regime.

As a test case, we have studied a water molecule in a standard
basis, comparing with the benchmark full configuration
interaction calculations of Bauschlicher and
Taylor\cite{bauschlicher}. In this work, exact results for the 
H$_2$O molecule within
a particular basis were compared with various approximate
approaches. (In the reference calculations and our work, the
innermost O orbital was ``frozen''.)
We have used the same basis for the calculations here.
In Fig. 1, we show DMRG results compared with Hartree Fock and
singles and doubles configuration interaction (SDCI), for
the molecule in its equilibrium geometry. Both
DMRG and SDCI results are variational. The DMRG results become
more accurate than the SDCI for $m \approx 70$.

\begin{figure}[ht]
\epsfxsize=3.0 in\centerline{\epsffile{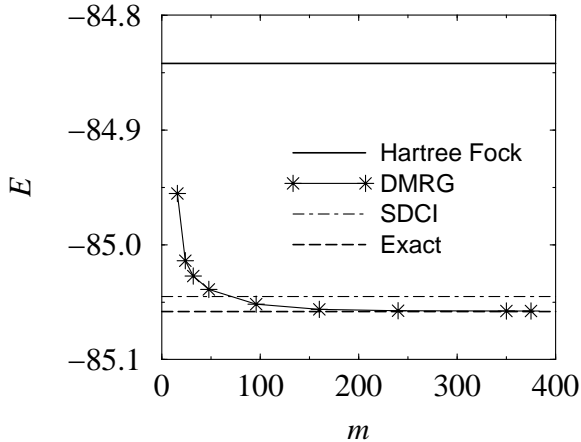}}
\caption{Ground state energy of a water molecule in a 25
orbital basis using various methods. DMRG results are as a
function of the number of states kept per block $m$; other
results, which have no $m$ dependence, are plotted as horizontal
lines. Energies are in Hartrees.
}
\end{figure}

In Fig. 2, we show results on an expanded scale for the same
system. We also compare with multireference
configuration interaction calculations (MRCI), and MRCI plus an
estimated correction (the Davidson correction) (MRCI+Q). MRCI
is variational, but MRCI+Q is not. We see that DMRG becomes
more accurate than MRCI for $m \approx 110$, and more accurate
than MRCI+Q for $m \approx 200$. The most accurate DMRG result
is off by 0.00024 Hartrees. The MRCI+Q results were the most
accurate of the approximate results reported by Bauschlicher and
Taylor, with an error of 0.0014.

\begin{figure}[ht]
\epsfxsize=3.0 in\centerline{\epsffile{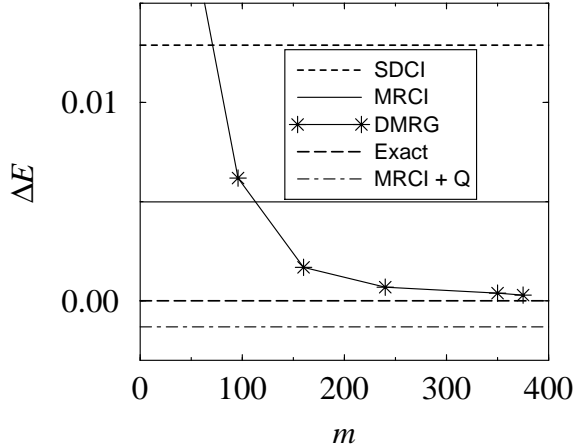}}
\caption{Difference between the exact ground state energy within the
given basis, and the approximate energy for various approaches. The
system and basis is the same as in Fig. 1. Energies are in
Hartrees.
}
\end{figure}

\begin{figure}[ht]
\epsfxsize=3.0 in\centerline{\epsffile{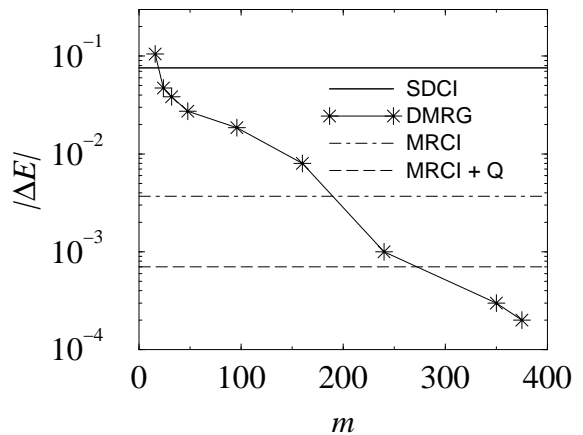}}
\caption{Absolute value of the difference between the exact ground state
energy within the given basis, and the approximate energy for various
approaches, measured in Hartrees. The
system is a water molecule with OH bonds stretched to twice their 
equilibrium length. The energy for MRCI+Q is below the exact result.
}
\end{figure}

Hartree Fock gives a reasonably adequate description of a water
molecule in its equilibrium geometry, but does not describe the
system well when one of the OH bonds is stretched significantly.
Many approximate approaches (such as SDCI) are strongly
dependent on the adequacy of the HF starting point. In order to
test the dependence of DMRG on the quality of HF, we have also
studied the water molecule with both OH bond lengths doubled.
This case was also studied by Bauschlicher and Taylor. In Fig. 3, 
we compare DMRG with SDCI, MRCI, and MRCI+Q. In this case, DMRG
performs better than SDCI starting much earlier, $m \approx 20$.
Both the MRCI and the DMRG results are largely unaffected by
the poor HF starting point. In this case, the most accurate DMRG 
result is off by 0.00019 Hartrees, while the MRCI+Q result is off by
0.00071 Hartrees. Interestingly, both results are more accurate
than for the equilibrium geometry. One can see that the convergence
of DMRG appears to be roughly exponential in $m$, which is also
usually the case in finite lattice model systems.

The numerical effort expended for these DMRG calculations was probably
more than for the other approximate methods considered above, but less
than for the full CI calculations. The point of our calculations
was not to present a fully developed technique, but to demonstrate
the potential of a new type of approach. DMRG is a versatile technique,
and we believe substantial improvements can be made over the
calculations described here. We will now discuss two general directions
we believe could be successful in improving the method.

An area for potential progress is in the choice of a single
particle basis. The Hartee Fock wavefunctions used here are
reasonable choices for a small molecule: this basis gives a
reasonable answer even if only one state per block is kept,
$m=1$.   Within the HF basis, the occupancy of orbitals tends to be
either almost 0 or almost 2, which helps the convergence of DMRG
as a function of $m$.  One could also try natural orbitals,
which have similar properties. However, both HF and natural
orbitals have a significant flaw: they are delocalized on a
large system.  Experience on lattice models suggests that DMRG
tends to be much more accurate with localized bases. An ideal
basis would be one that has occupancies close to 0 or 2, but
would also be as local as possible. The DMRG would thus seem to fit 
naturally into the local correlation approaches currently being developed\cite{pulay}.

Another area for potential progress is in the grouping of
similar orbitals into clusters. In the standard version of 
DMRG, blocks are formed for either the left or right
half of the system. However, it is also possible to form
blocks built out of clusters of orbitals which are strongly
coupled to each other. This sort of procedure was found to
be quite effective in an electron-phonon model, where a local
density matrix was used to reduce the size of the local phonon
space for each oscillator from up to 128 states to only 2 or 
3\cite{czhang}. Here,  one could form a cluster 
out of one ``occupied'' orbital plus a group ``virtual''
orbitals which are used to correlate the pair of electrons in
the occupied orbital. One would want to organize the basis set
into a set of such clusters,  with each cluster having one
occupied and a set of closely coupled virtual orbitals. A
density matrix would be used to form an accurate many body basis
for each of these clusters. Our preliminary calculations using
this approach suggest that fewer than 50 states would be sufficient
to describe such a cluster to millihartree accuracy, and even
using just a few states would probably be a substantial
improvement over Hartree Fock. (The Generalized Valence Bond
approach\cite{gvb} is closely related to this idea with $m=2$.)
A standard DMRG calculation could then procede using these
clusters as ``sites'', with the expectation that much higher
accuracy for a given $m$ would be obtained than in the above
orbital-by-orbital approach.  Alternatively, one might group
these clusters into superclusters, describing shells, atoms, or
even molecules.  In calculating the potential between two
molecules, for example, it would be particularly natural to use a
supercluster for each molecule. Note that efficient clustering
would require the use of localized orbitals, so that progress in
these two areas may be coupled.


We acknowledge support from the NSF under 
Grant No. DMR-9509945 (SRW), and the DOE/LDRD program at Los Alamos(RLM).


%

\begin{references}
\bibitem{details} In DMRG, for a 1D system with short ranged
interactions, the calculation time for a system of length
$L$, keeping $m$ states per block, is proportional to $m^3 L$.
The error in the energy of the ground state is believed
to vary as $\exp(-\alpha m)$ for systems with a gap to 
excited states. Since any finite system has gaps, we expect
to get exponential convergence asymptotically for any finite
system. However, the convergence using the infinite system
DMRG method in a gapless system is generally not exponential,
although it is usually still rather rapid.
\bibitem{notoned} Another difference is that this Hamiltonian is not
at all one dimensional, which decreases the accuracy of DMRG.
However, the correlations within an atom or small molecule are
quite different from in a model system: for example, the Hartree Fock
configuration is a much larger component of the ground state in
a molecule. In fact, numerical renormalization group methods
were originally developed to deal with a wide range of energy
scales, as found in molecular species.
\bibitem{twosite} Here, for simplicity, the two sites between
the two blocks in the standard superblock configuration are
assumed to be parts of their neighboring blocks.
\end{references}
\end{document}